\documentclass[pra,twocolumn,nofootinbib]{revtex4}

\usepackage{amssymb}
\usepackage{amsmath}
\usepackage{mathrsfs}
\usepackage{paralist}
\usepackage{graphicx}
\usepackage{xcolor}
\usepackage{fancyhdr}
\usepackage{xr}

\usepackage{soul}

\newcommand{\ii}{{\rm i}}
\newcommand{\e}{{\rm e}}

\begin{document}
%
% TITLE
%
    \title{Designing scattering-free isotropic index profiles using the phase-amplitude equations}
%
% AUTHORS
%
    \author{C. G. King}
    \author{S. A. R. Horsley}
    \author{T. G. Philbin}
    \affiliation{Department of Physics and Astronomy, University of Exeter, Stocker Road, Exeter, EX4 4QL}
%
%ABSTRACT
%
    \begin{abstract}
The Helmholtz equation can be written as coupled equations for the amplitude and phase. By considering spatial phase distributions corresponding to reflectionless wave propagation in the plane, and solving for the amplitude in terms of this phase, we have designed two-dimensional graded-index media which do not scatter light. We give two illustrative examples, the first of which is a periodic grating for which diffraction is completely suppressed at a single frequency at normal incidence to the periodicity. The second example is a medium which behaves as a 'beam-shifter' at a single frequency; acting to laterally shift a plane wave, or sufficiently wide beam, without reflection.
    \end{abstract}
    \maketitle
%
%START CONTENT
%
\section{Introduction:}
Wave propagation through inhomogeneous media cannot be solved analytically in most cases, even in one dimension. The space of possible media is also too large to be able to calculate reflection and transmission coefficients numerically in a representative sample of cases, particularly in higher dimensions where scattering can occur in various directions. Instead, mathematical techniques have been used to make progress, particularly with a view to designing non-scattering media. In one dimension, media whose graded-index susceptibility satisfies the spatial Kramers-Kronig relations are unidirectionally reflectionless for all angles of incidence~\cite{Horsley2016,Horsley2017} and remain reflectionless in two dimensions when their profiles are rescaled and translated along a second spatial coordinate. The analogous problem in higher dimensions is much harder to solve. Transformation optics~\cite{Pendry2006,Leonhardt2006} is a design procedure that removes reflections, but requires anisotropy and magnetic properties in general. In this work we devise an alternative mathematical method to design two dimensional scattering-free isotropic graded-index permittivity profiles based on mapping out the amplitude distribution, and hence permittivity profile, required to support a particular choice of phase in a lossless medium.
    \par
Designing reflectionless planar media (with material properties varying in only one dimension) is a difficult problem in itself and has been the consideration of considerable research in recent years. There are a small number of cases which can be solved exactly, such as the non-reflecting P{\"o}schl-Teller media~\cite{Epstein1930,Lekner2007} and Kay-Moses media~\cite{Kay1956}. The family of media whose susceptibility satisfy the spatial Kramers-Kronig relations~\cite{Horsley2015,Horsley2016} have been used to design disordered permittivity profiles exhibiting perfect transmission~\cite{King2017,Makris2017} and perfectly absorbing media~\cite{King2017(2)}. Experimental realisations of near perfect absorbers based on these media have been carried out in~\cite{Jiang2017,Ye2017}.
    \par
Finding reflectionless media that induce some specified change in the wave is, unsurprisingly, a more difficult problem than the one-dimensional analogue. With the extra complications, however, comes a greater range of practical possibilities, such as beam bending, shifting or focussing, as well as cloaking. Ray tracing (see for example~\cite{Leonhardt}) can be used to design inhomogeneous refractive index profiles that guide the wave's energy in some specified way. For example, radial index profiles, such as the Luneburg lens~\cite{Luneburg1944}, can be used to focus light from a plane wave to a single point. However, such an approach relies on the validity of the geometrical optics approximation, which will, for example, break down near the focus of the lens, and also ignores the phenomenon of reflection. By considering the exact wave problem, we are able to bypass such difficulties enabling a greater control of the sort of frequencies our media can suppress diffraction to. Transformation optics~\cite{Pendry2006,Leonhardt2006} has been at the forefront of recent developments; in particular leading to the practical possibility of cloaking~\cite{Kundtz2010,Schurig2006}. Instead of designing materials through coordinate transformations, our approach is to put the exact phase rays at the forefront and work out what sort of amplitude distribution and permittivity profile is required to guide a wave through an object without scattering. This 'reverse engineering' approach of working out the material properties leading to a pre-specified wave solution has been considered in two dimensions before for long range materials~\cite{Vial2017,Liu2017}. The non-magnetic materials designed in~\cite{Vial2017} are only designed to work approximately, however, since the phase gradient specified does not have a vanishing curl. We explore what non-magnetic materials based on exact reflectionless solutions can be found. We give two applications of our method. Firstly, we design media, periodic in one direction, which transmit perfectly without reflection or diffraction for waves incident perpendicular to the periodicity, for a surprisingly large bandwidth, Secondly, we design two dimensional reflectionless beam-shifters for a single frequency and angle of incidence.
    \par
\begin{figure}[ht!]
        \begin{center}
	\includegraphics[width=\linewidth]{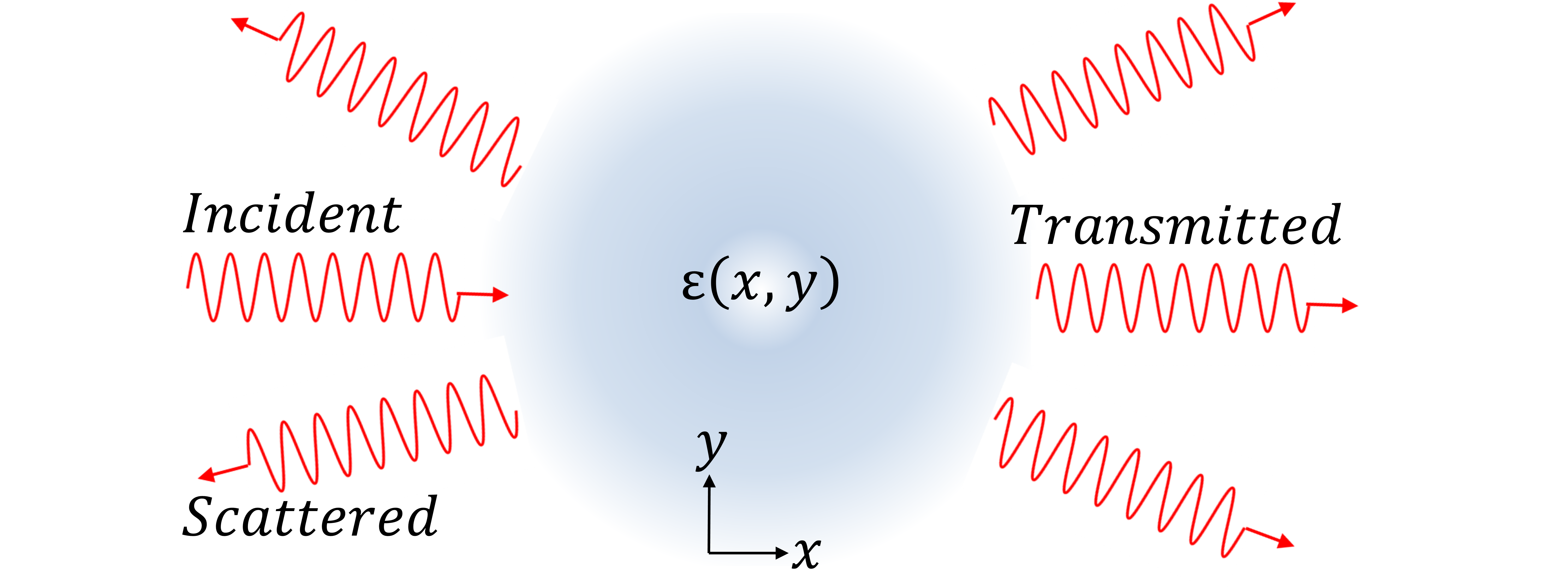}
        \caption{A wave of wave vector \(\textbf{k}=(k_{x},0,0)\) is incident from free space onto a planar medium inhomogeneous in the \((x,y)\)-axis characterised by a real-valued graded-index permittivity \(\epsilon(x,y)\), where \(\epsilon\to1\) as \(x\to\pm\infty\). The magnitude and direction of the scattering from the medium will depend on the spatial variation of $\epsilon$.\label{model}}
        \end{center}
    \end{figure}

Consider the 2-dimensional situation sketched in figure~\ref{model} for propagation of electromagnetic waves through a slab of material embedded in free space.
The out-of-plane component of the electric field corresponding to a monochromatic Transverse Electric (TE) polarised wave of frequency \(\omega\) incident upon a medium with permittivity \(\epsilon\) satisfies the Helmholtz equation
    \begin{equation}
        \left[\nabla^{2}+k_{0}^{2}\epsilon\right]\varphi=0.\label{1}
    \end{equation}
where \(k_{0}=\omega/c\) is the wave number. When $\epsilon$ is allowed to be an arbitrary function of position, there are only a few exactly solvable cases. Instead of attempting to solve directly, it is convenient to rewrite the solution in terms of its positive amplitude \(A\) and real-valued phase \(S\) as \(\phi=A(x,y)\e^{\ii k_{0}S(x,y)}\). Upon substitution back into (\ref{1}) and separation into real and imaginary parts, we obtain the following equations relating the amplitude and phase to the permittivity, which we call the phase-amplitude equations:
    \begin{equation}
    \begin{split}
        \epsilon&=(\nabla S)^{2}-\frac{\nabla^{2}A}{k_{0}^{2}A}\\
        0&=\nabla\cdot(A^{2}\nabla S),\label{2}
    \end{split}
    \end{equation}
where we assume that the permittivity \(\epsilon\) is real valued, such that the second equation depends only on the amplitude and phase of the wave. These two equations describe the intricate relationship between amplitude and phase and are key to determining the required distribution of amplitude (and hence permittivity) needed to guide rays in a desired way. Reference~\cite{Philbin2014} explores special cases in which geometrical optics gives the exact solution to (\ref{2}) i.e. when the 'quantum potential' term \(\frac{\nabla^{2}A}{A}\) vanishes. Instead, here we solve the second equation, and use the first to obtain the corresponding permittivity. The divergence free quantity \(A^{2}\nabla S\) is exactly proportional to the time-averaged Poynting vector (more precisely \(\textbf{S}=\frac{1}{2\mu_{0}}A^{2}\nabla S\)), and this equation is simply an energy conservation equation expressing the assumption that no current sources are present in the medium. For any region of the \((x,y)\) plane, the rate of energy flow into the region via the electromagnetic field must equal the rate of energy leaving the region-: it is precisely this simple principle that we exploit in our design of perfectly transmitting two-dimensional lossless media.
    \par
For propagation in one dimension, where the energy flow can only be forward or backward propagating, the second phase-amplitude equation in (\ref{2}) (the energy conservation condition) takes a particularly simple form
    \begin{equation}
        \frac{d}{dx}\left(A^{2}\frac{dS}{dx}\right)=0,\label{14}
    \end{equation}
which can immediately be integrated up to give
    \begin{equation}
        A=\frac{A_{0}}{\sqrt{\frac{dS}{dx}}}.\label{15}
    \end{equation}
By choosing a phase distribution \(S(x)\) corresponding to a plane wave of unidirectional propagation asymptotically (e.g. \(S\sim x\) as \(x\to\pm\infty\)), the corresponding amplitude is determined from (\ref{15}) and then a reflectionless permittivity profile can be found from the first equation of (\ref{2}). Alternatively, it is common to take the geometrical optics limit \(k_{0}\gg|\nabla\epsilon|/\epsilon^{3/2}\), where the remaining phase-amplitude equation is simply the eikonal equation \(\epsilon=\left(\frac{dS}{dx}\right)^{2}\) and the WKB approximations are found~\cite{Heading}.
    \par
Extending this to higher dimensions, where the energy flow can be in a number of directions, is non-trivial. With the aim of finding non-scattering media, we choose a phase distribution \(S(x,y)\) corresponding to a plane wave of unidirectional propagation asymptotically (i.e. still imposing \(S\sim x\) as \(x\to\pm\infty\)) and then use equations (\ref{2}) to find a corresponding amplitude and permittivity that permits such directional control of the wavefronts.

\section{The Characteristic method:}
\begin{figure}[ht!]
        \begin{center}
	\includegraphics[width=\linewidth]{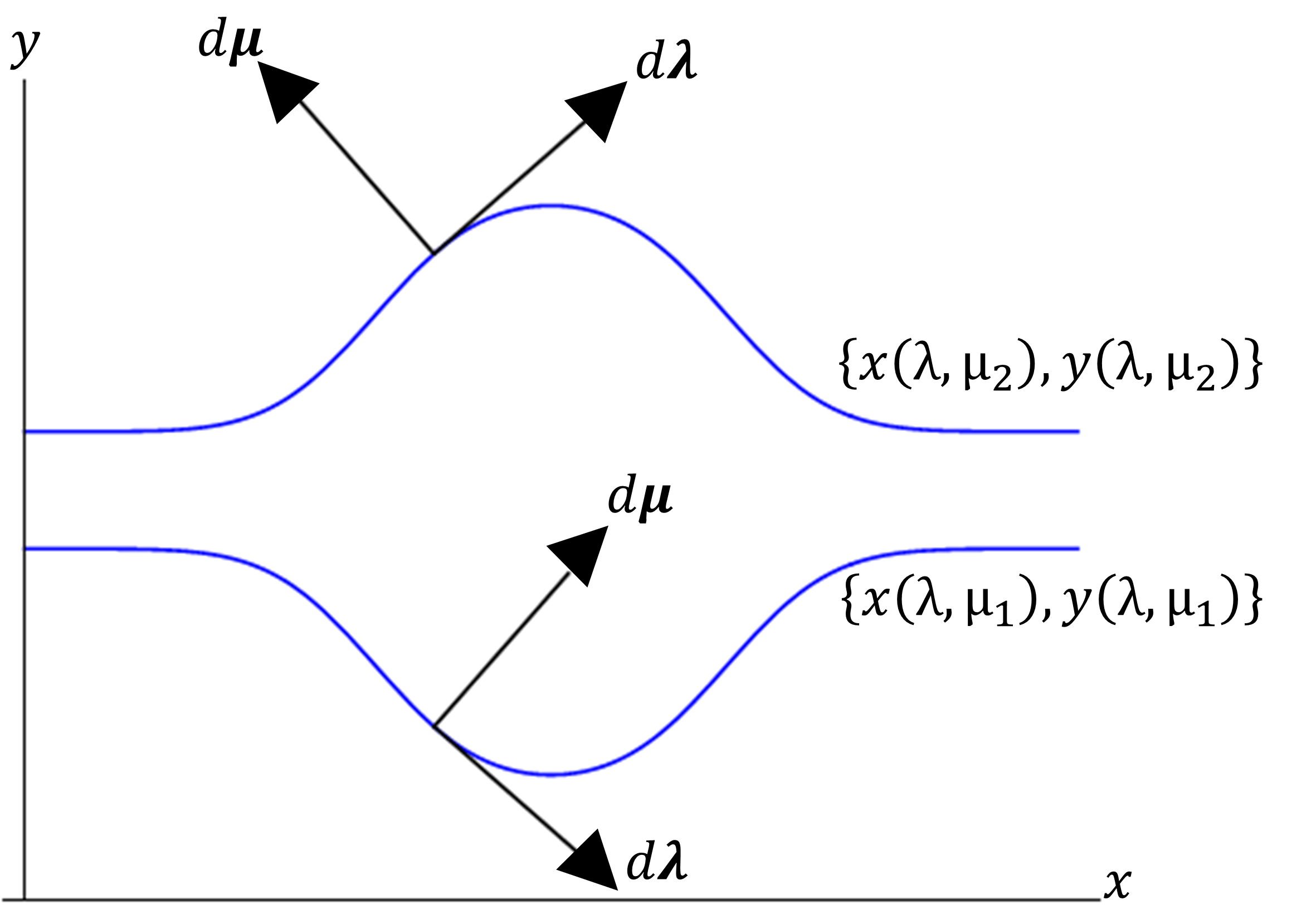}
        \caption{Plots of two of the rays%in (a) \((x,y)\) space, and (b) \((\lambda,\mu)\) space. \((\lambda,\mu)\) defines a coordinate system in which the rays are straight lines of constant \(\mu\)%
. \(\lambda\) parameterises the curve describing each ray whilst \(\mu\) parameterises the different rays.%In \((\lambda,\mu)\) space, the equation for amplitude becomes a simple ODE.%
\label{characteristicmethod}}
        \end{center}
    \end{figure}

The second phase-amplitude equation in (\ref{2}) can be written as
    \begin{equation}
        \frac{\partial S}{\partial x}\frac{\partial A}{\partial x}+\frac{\partial S}{\partial y}\frac{\partial A}{\partial y}=-\frac{A}{2}\nabla^{2}S.\label{3}
    \end{equation}
This is now of the form for which the method of characteristics may be applied (see, for example, \cite{Courant1989} for a discussion of this method). Therefore the following set of equations should be solved simultaneously:
    \begin{equation}
    \begin{split}
        \frac{dx}{d\lambda}&=\frac{\partial S}{\partial x}\\
        \frac{dy}{d\lambda}&=\frac{\partial S}{\partial y}\\
        \frac{dA}{d\lambda}&=-\frac{A}{2}\nabla^{2}S.\label{4}
    \end{split}
    \end{equation}
Together with suitable boundary conditions, the resulting parametric solution will map out a surface in \((x,y,A)\) space. The first two equations decouple from the third and can be numerically solved to map out the rays (or characteristics) in the \((x,y)\) plane with the parameter \(\lambda\) parameterising each ray (as can be seen by taking their ratio \(\frac{dy}{dx}=\frac{\partial_{y}S}{\partial_{x}S}\)). The different rays are parameterised by a different parameter, \(\mu\) say, depending on the specific form of the boundary condition. This is similar to the method of Transformation optics described in \cite{Leonhardt2006} in that we start with a coordinate system (\(\lambda\), \(\mu\)) in which the rays are globally straight and parallel as is supported by free space, and then find a transformation (\(x(\lambda,\mu)\),\(y(\lambda,\mu)\)) to a new coordinate system in which the rays behave in a particular desired fashion. However, we do not confine our mappings to be conformal (i.e. angle preserving). For example, we can just as easily rescale the parameters \(\lambda\) and \(\mu\) independently to our convenience. In fact it is computationally quicker to rescale \(\lambda\) to correspond to actual distance along the ray when mapping out the rays numerically, rather than the phase of the wave (so curves of constant \(\lambda\) need not be the phase fronts). The situation is described visually in figure~\ref{characteristicmethod}.
    \par
In general, given a boundary condition for the amplitude along a curve, \(C\) in the \((x,y)\) plane, there will be a unique solution in the region of the \((x,y)\) plane spanned by the rays passing through \(C\). In particular, we can impose a uniform amplitude along a vertical line \(x=\)constant on the left (incident) side of the medium. Together with a  suitable choice of phase \(S\), this will correspond to a right propagating plane wave incident without reflection. As an example of this method, we describe how periodic media can be designed to have no diffraction, although we envisage the method being useful for the design of other sorts of non-scattering media in two and three dimensions, such as beam-benders and lenses.

\section{Reflection and Transmission coefficients for a periodic medium:}
\begin{figure}[ht!]
        \begin{center}
	\includegraphics[width=\linewidth]{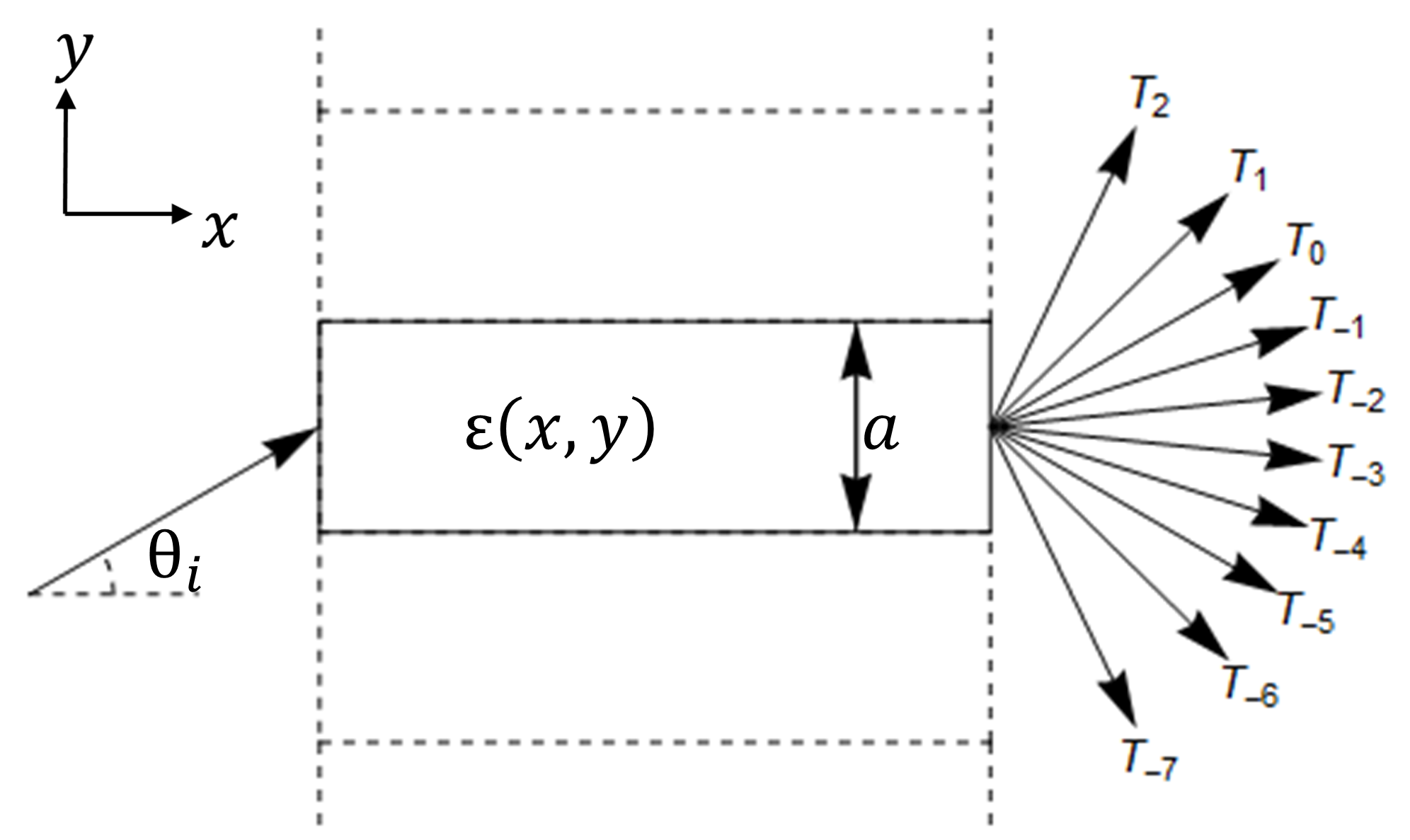}
        \caption{A wave incident from the left upon a permittivity profile \(\epsilon(x,y)\) periodic in the \(y\) direction with period \(a\). The resulting diffraction pattern consists of a superposition of waves reflected and transmitted at angles \(\theta_{n}\) given by equation (\ref{5}) with intensities given by equation (\ref{10}) (the reflected waves aren't shown in this diagram to avoid cluttering).\label{angles}}
        \end{center}
    \end{figure}

To motivate the application of our method to designing diffractionless gratings, we now review diffraction theory. Consider a plane wave propagating in the positive \(x\) direction impinging on a medium periodic in the \(y\) direction, with periodicity \(a=2\pi/k_{g}\) and sitting in free space: \(\epsilon\to1\) as \(x\to\pm\infty\). Such a periodic medium will typically produce a diffraction pattern. Relative to an angle of incidence \(\theta_{i}\) with the positive \(x\) axis, the possible angles for waves to scatter away from the medium are
    \begin{equation}
        \text{sin}\theta_{n}=\text{sin}\theta_{i}+\frac{nk_{g}}{k_{0}}.\label{5}
    \end{equation}
The periodicity of the profile ensures that the field outside the medium can be naturally written as a Fourier series of reflected and transmitted waves propagating in various directions:
    \begin{equation}
        \varphi=
    \begin{cases}
        \e^{\ii\textbf{k}_{0}\cdot\textbf{x}}+\sum_{n=-\infty}^{\infty}\varphi_{r,n}\e^{\ii\textbf{k}_{r,n}\cdot\textbf{x}}&\qquad x\sim-\infty\\
        \sum_{n=-\infty}^{\infty}\varphi_{t,n}\e^{\ii\textbf{k}_{t,n}\cdot\textbf{x}}&\qquad x\sim+\infty.\label{6}
    \end{cases}
    \end{equation}
where
    \begin{equation}
    \begin{split}
        \textbf{k}_{0}&=\sqrt{k_{0}^{2}-k_{y}^{2}}\hat{\textbf{x}}+k_{y}\hat{\textbf{y}}\\
        \textbf{k}_{r,n}&=-\sqrt{k_{0}^{2}-(k_{y}+nk_{g})^{2}}\hat{\textbf{x}}+(k_{y}+nk_{g})\hat{\textbf{y}}\\
        \textbf{k}_{t,n}&=\sqrt{k_{0}^{2}-(k_{y}+nk_{g})^{2}}\hat{\textbf{x}}+(k_{y}+nk_{g})\hat{\textbf{y}}.\label{7}
    \end{split}
    \end{equation}
It is constructive to restrict ourselves to consideration of the wave behaviour in a single 'unit cell' of the periodic medium. More specifically, consider the region of the \((x,y)\) plane bounded by rays separated in the \(y\) direction by a distance \(a\). There can be no flow of energy in or out of such a region when averaged over time, by construction. Therefore, any net energy flow into the medium from \(x=-\infty\) must equate to the energy flow exiting the medium at \(x=\infty\). For example
    \begin{equation}
        \int_{-a/2}^{a/2}dy\textbf{S}\cdot\hat{\textbf{x}}|_{x\to-\infty}=\int_{-a/2}^{a/2}dy\textbf{S}\cdot\hat{\textbf{x}}|_{x\to+\infty}.\label{8}
    \end{equation}
Upon substitution of the periodic field (\ref{6}) into the energy conservation equation (\ref{8}), we obtain the following relationship:
    \begin{equation}
        \sum_{n=M}^{N}(R_{n}+T_{n})=1,\label{9}
    \end{equation}
where \(R_{n}\) and \(T_{n}\) are the n\(^{th}\) reflection and transmission coefficients, respectively, describing the power going into the n\(^{th}\) order reflected and transmitted modes. They are given by
    \begin{equation}
    \begin{split}
        R_{n}&=\frac{\sqrt{k_{0}^{2}-(k_{y}+nk_{g})^{2}}}{\sqrt{k_{0}^{2}-k_{y}^{2}}}|\varphi_{r,n}|^{2}\\
        T_{n}&=\frac{\sqrt{k_{0}^{2}-(k_{y}+nk_{g})^{2}}}{\sqrt{k_{0}^{2}-k_{y}^{2}}}|\varphi_{t,n}|^{2}.\label{10}
    \end{split}
    \end{equation}
where the coefficients in the solution, \(\varphi_{r,n}\) and \(\varphi_{t,n}\), are calculated as the Fourier components of (\ref{6}) and the sum is taken over all propagating modes (\(|k_{y}+nk_{g}|<k_{0}\)). The situation is illustrated in figure~\ref{angles}.

\section{The Non-diffracting Grating:}
In this section we apply the earlier method of characteristics to the problem of designing a diffraction grating that doesn't diffract. This means that all of the wave's energy is carried by the zero order transmitted mode, with the rays emerging undeviated. Strong diffraction is usually computed numerically; however we are giving a design procedure by which one can specify where the diffraction is zero. There has been some recent work on diffraction from periodic structures exhibiting Parity-Time (PT) symmetry, illustrating the asymmetry in the diffracted fields~\cite{Makris2008,Ruter2010}. Also, the reflectivity and transmissivity of discrete periodic gratings has been studied as wavelength and incidence angle is varied in~\cite{Fitio2005} with a view to improving diffraction efficiency. In particular, efficiency of diffraction to the first order reflected mode has been improved whilst suppressing the zero order reflected mode using plasmonic metasurfaces~\cite{Zhang2016}. In our example we show how to perform the polar opposite function; namely to improve the efficiency of energy going into the zero order transmitted mode by minimising the energy going into all other modes. However, our theory can be used to manipulate the diffraction from a periodic structure in a quite arbitrary way.
    \par
Consider designing a profile for which diffraction is suppressed for a particular wavenumber \(\kappa_{0}\) at and angle \(\theta_{i}\) to the \(x\) axis. Then, for a right propagating plane wave with perfect transmission without reflection, the phase should asymptotically satisfy \(S\sim\text{cos}\theta_{i}x+\text{sin}\theta_{i}y\) as \(x\to\pm\infty\), and any distortion in the rays should be confined within the medium. To this end we make the following choice of phase. In the region \(-\frac{a}{2}<y<\frac{a}{2}\), let
    \begin{equation}
    \begin{split}
        S&=\text{cos}\theta_{i}x+\text{sin}\theta_{i}y+b\text{erf}\left(\frac{x}{c}\right)\\
        &\quad+\alpha x\e^{-\left(\frac{x}{d}\right)^{2}}\left[1+\text{erf}\left(\frac{\frac{a}{4}+y}{h}\right)\text{erf}\left(\frac{\frac{a}{4}-y}{h}\right)\right],\label{11}
    \end{split}
    \end{equation}
where erf\((z)=\frac{2}{\sqrt{\pi}}\int_{0}^{z}\e^{-\tilde{z}^{2}}d\tilde{z}\) is the error function, which switches smoothly from \(-1\) to \(+1\) with increasing argument. This is then repeated periodically up and down the \(y\) axis. For simplicity, propagation at normal incidence to the periodicity (\(\theta_{i}=0\)) will be studied, with the understanding that the method exactly extends to non-normal incidence. Due to the \(y\) dependence of the phase, there will be a slight discontinuity in the phase gradient across the edges of the unit cell. However, the exponential decay of the error functions ensures that this jump is exponentially small and therefore will have a negligible effect in the calculation of the reflection and transmission coefficients (indeed with the parameter values used in the example plotted in this section, the jump is five orders of magnitude smaller than the maximal value of the phase gradient).
    \par
The choice of phase given in (\ref{11}) is of course by no means a unique choice and it is therefore sensible to motivate this particular choice. Bearing in mind the aim is for an invisible periodic profile, we need a phase corresponding to a unidirectional wave either side of the medium and thus we choose a phase with leading order asymptotic behaviour \(S\sim x\) as \(x\to\pm\infty\) corresponding to a right propagating wave without lateral scattering. The \(x\) dependent error function determines the phase shift \(\e^{2\ii k_{0}b}\) of the wave upon propagation through the medium with the parameter \(c\) being a measure of the scale upon which this shift occurs. This term has been included to ensure that the relative permittivity remains above unity so that the medium could be fabricated out of normal dielectric media. The final term encodes the transverse dependence of the phase, whose strength diminishes exponentially away from the \(y\) axis. This term has been multiplied by \(x\) to ensure an odd symmetry dependence of the phase on this coordinate. This guarantees that the spacing between the rays before and after propagation through the medium remains the same i.e. the rays don't 'bunch up' as a result of transmitting through the medium.
    \par
\begin{figure}[ht!]
        \begin{center}
	\includegraphics[width=\linewidth]{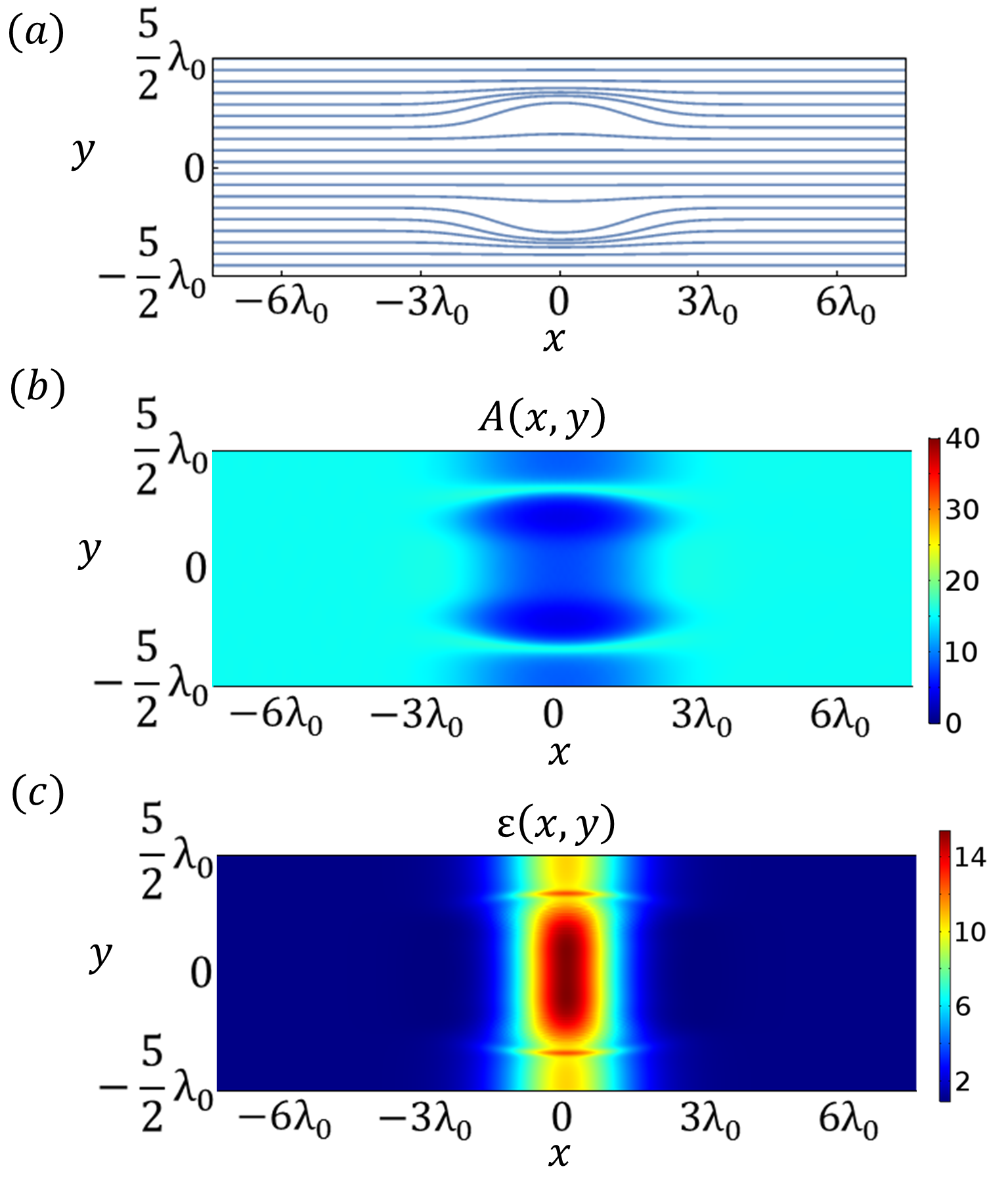}
        \caption{(a) The rays corresponding to the phase distribution given in (\ref{11}) in a 'unit cell' where period, \(a=5\lambda_{0}\), \(b=10\lambda_{0}/\pi\), \(c=d=5\lambda_{0}/\pi\), \(h=5\lambda_{0}/4\pi\) and \(\alpha=1/3\). (b) The corresponding amplitude resulting from solving the characteristic equations (\ref{4}). (c) The corresponding permittivity profile as determined from the first equation in (\ref{2}) corresponding to \(k_{0}=\kappa_{0}\).\label{rays}}
        \end{center}
    \end{figure}

Given a particular choice of wavelength \(\lambda_{0}=2\pi/\kappa_{0}\) for the incident plane wave, We have chosen a combination of parameters in (\ref{11}) which ensures that diffraction should be possible (indeed the first four diffracted modes should be visible) but completely vanishes. The characteristic method can then be used to numerically find the exact rays corresponding to this choice of phase, as shown in figure~\ref{rays}(a).
Notice that the dependence of the medium transverse to the direction of propagation acts to distort the rays (and correspondingly, the phasefronts). However, upon propagation through the medium, the rays respace evenly again. This method is exact and does not rely on the approximation of geometrical optics, so a plane wave incident upon the medium will emerge as a plane wave, if the appropriate boundary condition for the amplitude, \(A\to1\) as \(x\to-\infty\), is applied. The final equation of (\ref{4}) is solved subject to this boundary condition and the resulting amplitude is plotted in figure~\ref{rays}(b). The uniformity of the amplitude either side of the medium implies that a monochromatic plane wave propagates through the medium without scattering (whether in the form of reflection or diffraction). The corresponding permittivity profile is plotted in figure~\ref{rays}(c). As expected, the permittivity approaches that of free space as \(x\to\pm\infty\) with a range of unity up to around 15 in the medium in this particular example. We have chosen a combination of parameters such that the absence of diffraction is a surprising result, but not so high that the absence of reflection can be put down to being in the geometrical optics limit, since the spatial variation of the permittivity is on the order of a wavelength. In the geometrical optics limit, the permittivity is simply given by the eikonal equation \(\epsilon=|\nabla S|^{2}\). However, in the exact wave optics, to which the characteristic method applies, the permittivity includes the quantum potential term \(-\nabla^{2}A/k_{0}^{2}A\) and this term is significant (dominant) when the permittivity varies on a scale on the order of (much shorter than) the wavelength. The characteristic method can be used for any of these regimes.
    \par
\begin{figure}[ht!]
        \begin{center}
	\includegraphics[width=\linewidth]{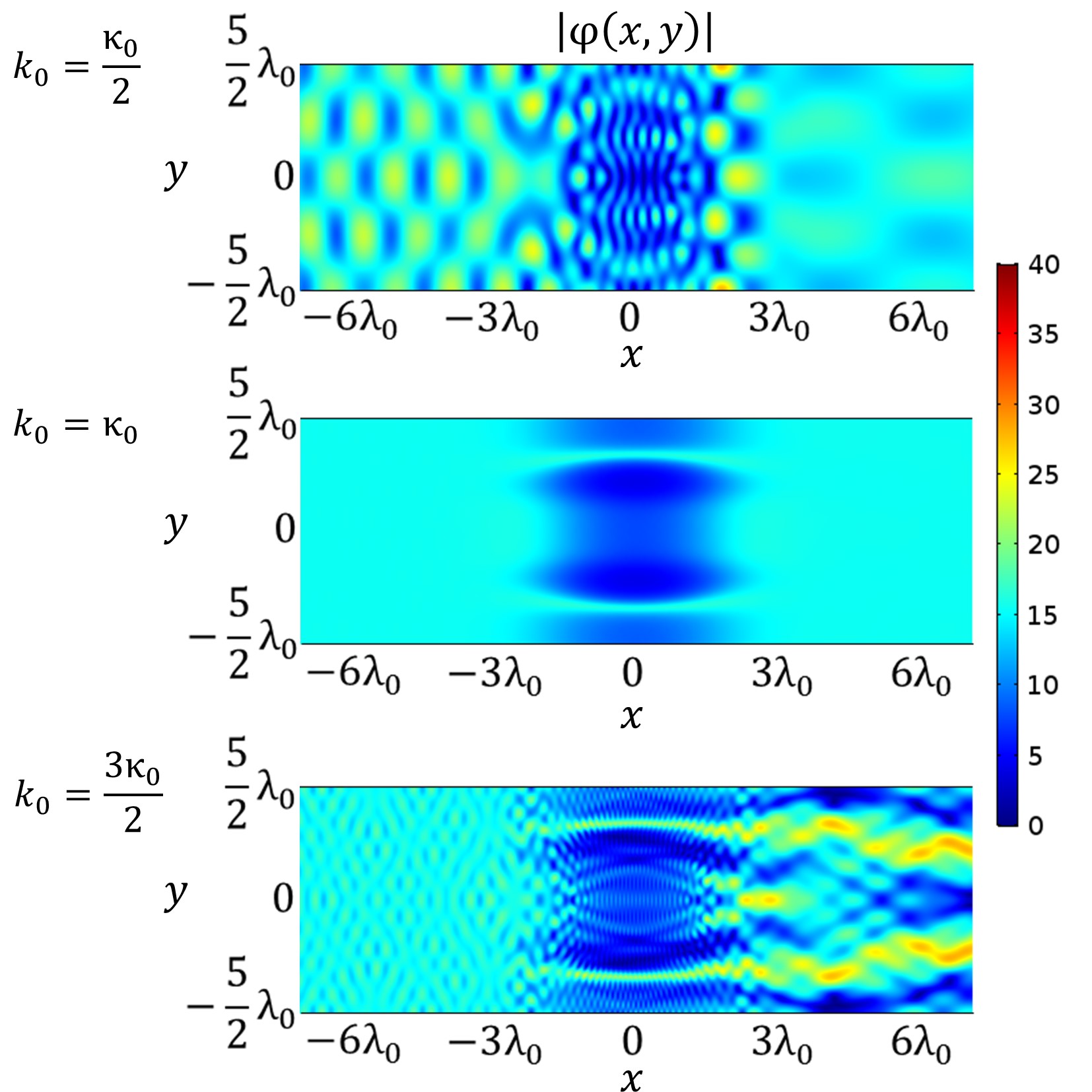}
        \caption{The electric field norm corresponding to a plane wave propagating in the positive \(x\) direction through the permittivity profile of figure~\ref{rays}(c) at four different wave numbers, simulated  using Comsol Multiphysics~\cite{Comsol}. As expected the field amplitude is uniform at the wavenumber \(k_{0}=\kappa_{0}\) designed to give no scattering whereas diffraction is visible at other wave numbers.\label{diffractionpattern}}
        \end{center}
    \end{figure}
\begin{figure}[ht!]
        \begin{center}
	\includegraphics[width=\linewidth]{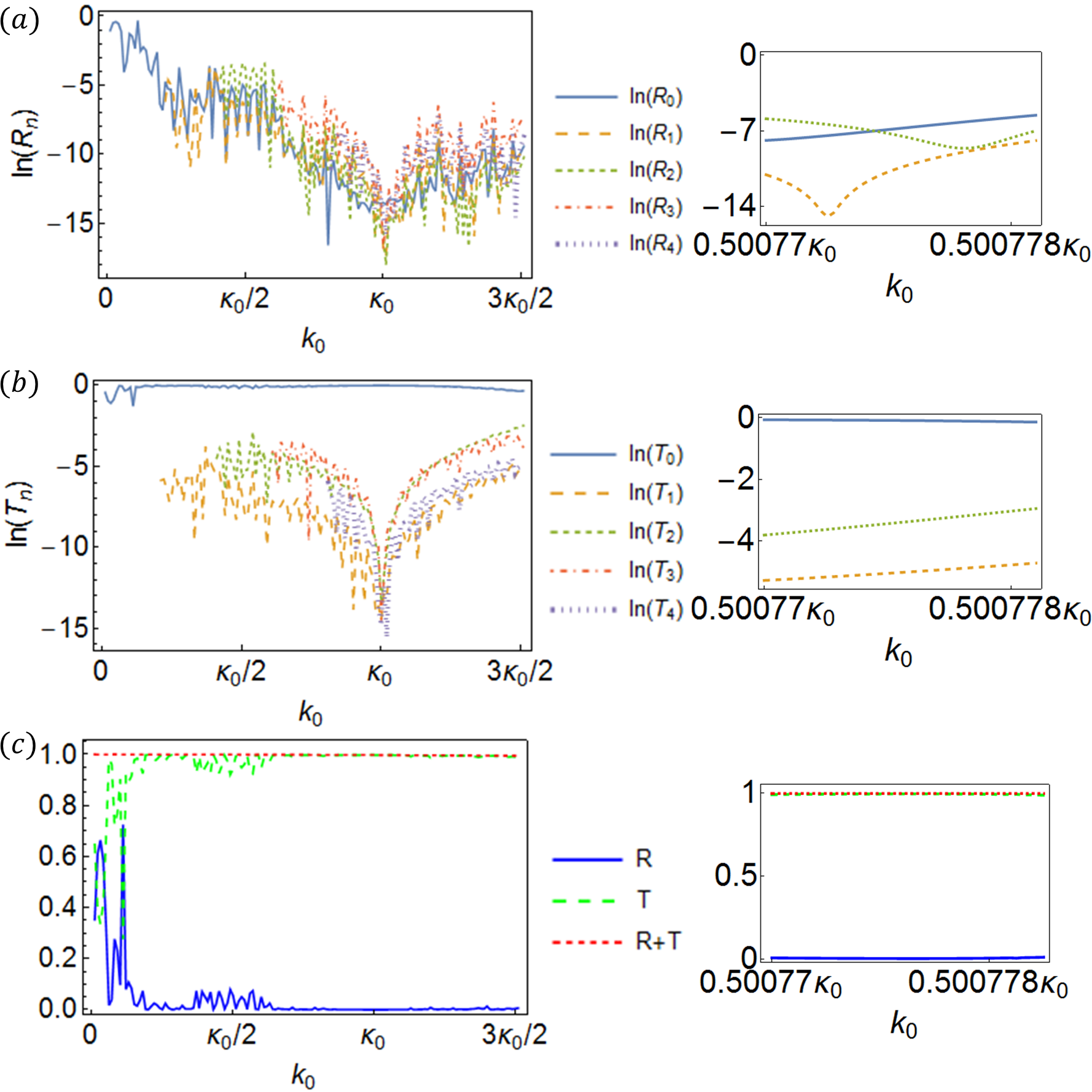}
        \caption{The natural logarithm of the (a) reflected and (b) transmitted intensities of the first five non-negative diffracted modes that a right propagating plane wave impinging on the permittivity profile of figure~\ref{rays}(c) at normal incidence scatters into as a function of wavenumber. (c) The total reflected (\(R\)) and transmitted (\(T\)) intensities and their sum. Over 99\(\%\) of the wave energy ends up in the \(T_{0}\) mode in the band \(0.92\kappa_{0}<k_{0}<1.08\kappa_{0}\). Thus diffraction is suppressed in quite a broadband region. Only positive Fourier components are plotted because the reflectional symmetry in the profile \(\epsilon(x,-y)=\epsilon(x,y)\) ensures that negative mode intensities match the corresponding positive mode intensities at normal incidence. Despite the curves appearing to lack the smoothness over the domain \(k_{0}\in[0,3\kappa_{0}/2]\) that would be expected of a continuous change of the wavenumber, the graphs are in fact smooth, as can be seen from the close-ups to the right.\label{varywavenumber}}
        \end{center}
    \end{figure}

The design process of such a non-scattering medium is such that it is only expected to function for an incident plane wave of frequency \(\omega=c\kappa_{0}\) for normal incidence. For other frequencies there is no reason not to expect a large amount of scattering in the form of both reflected and transmitted diffracted waves. We have investigated the effect of sending in different frequencies of radiation through the permittivity profile. Some plots of the electric field norm are shown in figure~\ref{diffractionpattern}.
Except at the wavenumber for which the medium is designed to be non-scattering, we see intricate diffraction patterns on both the incident and transmitted sides of the medium, with the fineness of the pattern being on the order of the wavelength. Variations in the electric field norm on either side of the medium indicate interference between plane waves propagating in different directions. Upon calculation of the reflected and transmitted intensities of these waves (\ref{10}), plots of the intensities as a function of wavenumber can be obtained and are shown for this example in figure~\ref{varywavenumber}.
For \(k_{0}<\kappa_{0}/5\), the wavelength is longer than the periodicity, diffraction is not possible and so only the zero order modes corresponding to lateral transmission and reflection are possible. As \(k_{0}\) is increased beyond \(\kappa_{0}/5\), diffraction is expected and energy is carried by the first order modes via both reflection and transmission. As \(k_{0}\) is increased further, higher order modes can carry energy. It appears from figure~\ref{varywavenumber}(c) that reflection is well suppressed beyond around \(k_{0}=0.6\kappa_{0}\). However, when we plot out the individual mode intensities on a log scale in (a) and (b), it can be seen that diffraction is strongly suppressed only in a smaller band around \(k_{0}=\kappa_{0}\) with some more significant diffraction outside this band. Also, since these graphs compare intensities (which are proportional to the square of the wave amplitude), diffraction is more prominent in the wave amplitude plots in figure~\ref{diffractionpattern} outside the smaller band. Whilst it is clear from figure~\ref{varywavenumber} that the intensities fluctuate very rapidly as the wavenumber is altered through different sharp resonances, there is a noticeable band dip (broader than the sharp resonances) in all but the zero order transmitted mode (the unscattered mode) around the wavenumber \(k_{0}=\kappa_{0}\) at which the structure is designed to be reflectionless and perfectly transmitting. Thus, it is perfectly feasible for a long Gaussian pulse (for example) consisting predominantly of a small range of frequencies close to \(\kappa_{0}c\) to scatter negligibly.
    \par
\begin{figure}[ht!]
        \begin{center}
	\includegraphics[width=\linewidth]{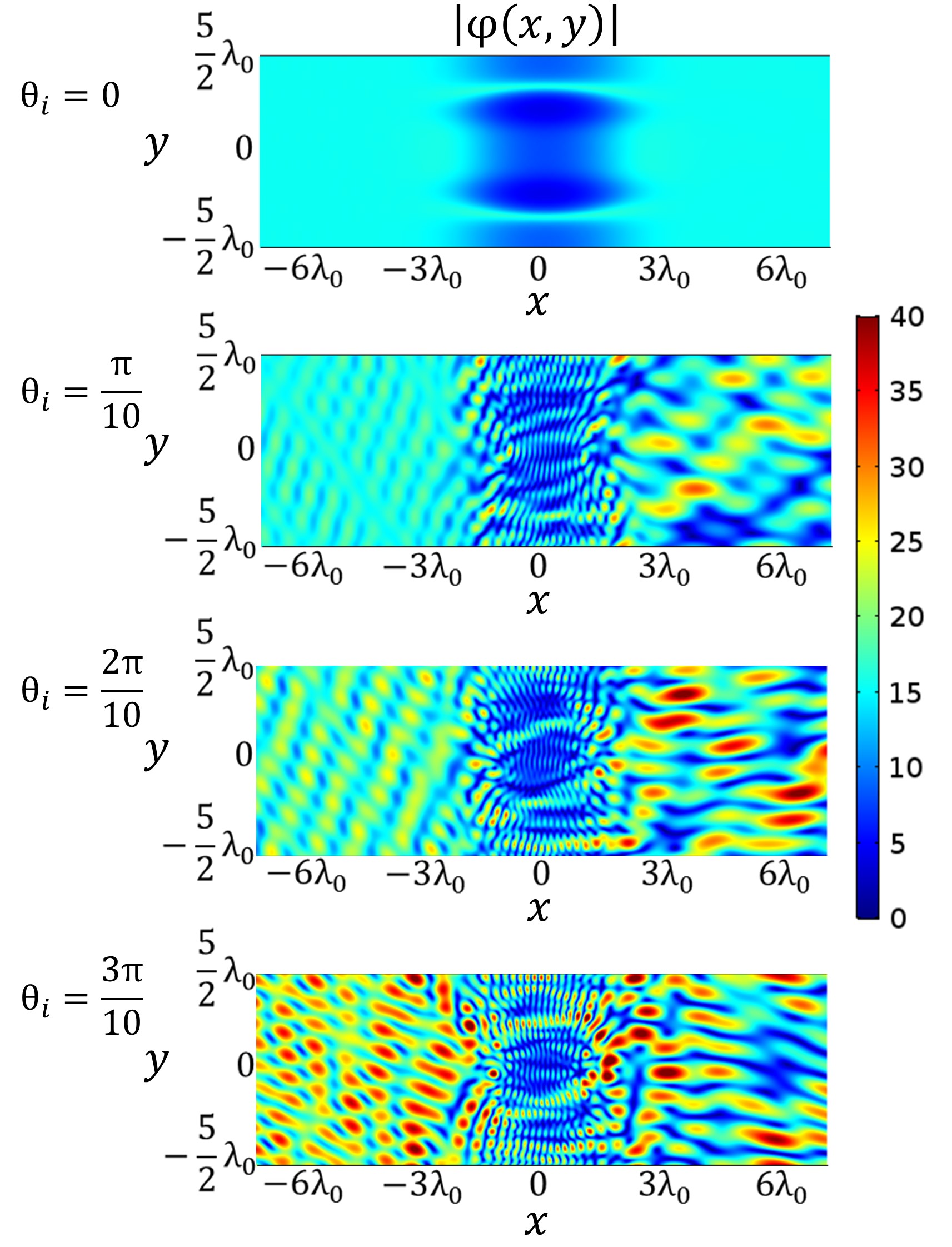}
        \caption{The electric field norm corresponding to a plane wave propagating at various angles \(\theta_{i}\) to the positive \(x\) axis through the permittivity profile of figure~\ref{rays}(c) at wavenumber \(k_{0}=\kappa_{0}\), simulated using Comsol Multiphysics~\cite{Comsol}. For angles off normal incidence, diffraction is again visible.\label{diffractionpatternvaryangle}}
        \end{center}
    \end{figure}
\begin{figure}[ht!]
        \begin{center}
	\includegraphics[width=\linewidth]{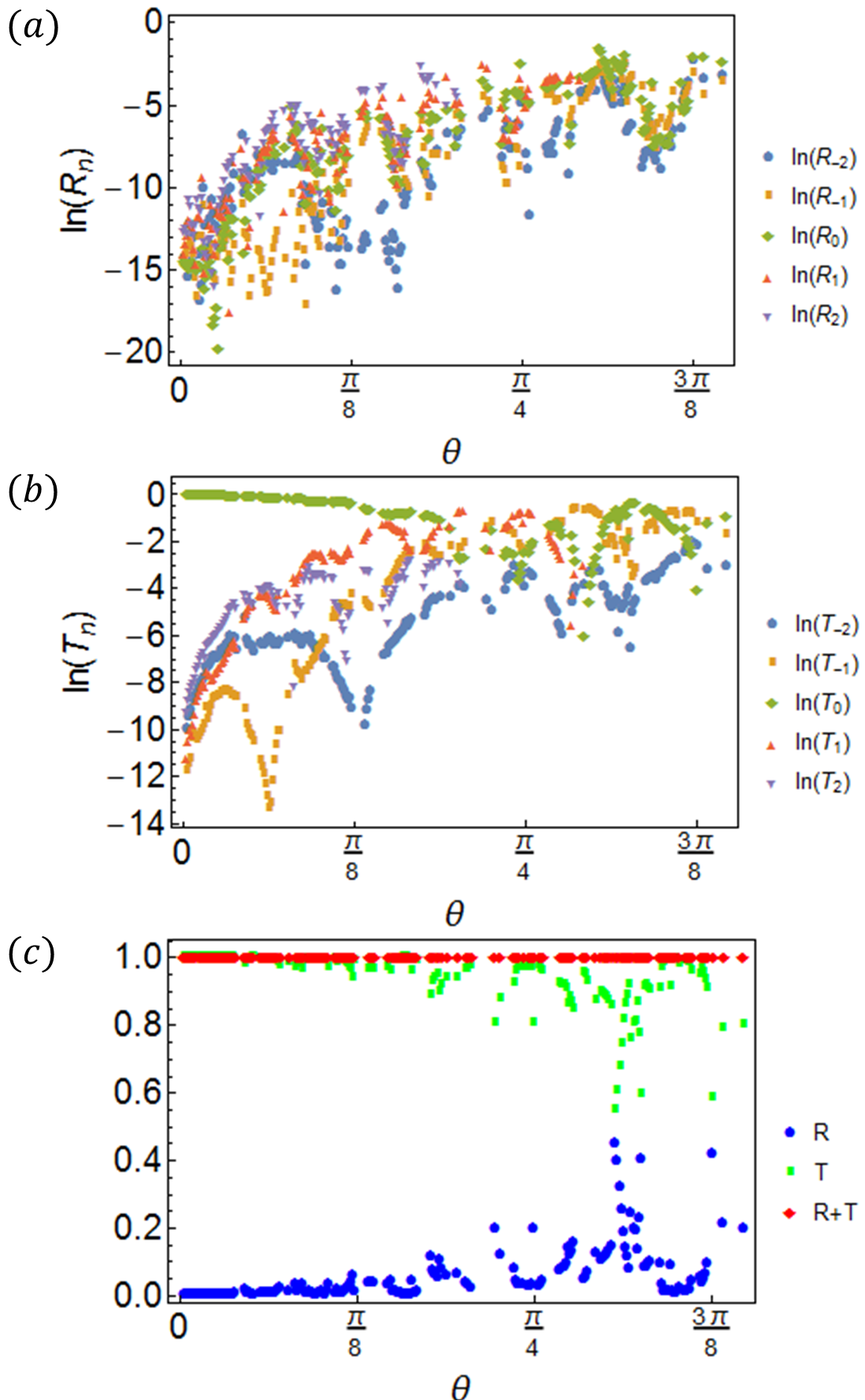}
        \caption{The natural logarithm of the (a) reflected and (b) transmitted intensities of the five middle diffracted modes that a plane wave of wavenumber \(k_{0}=\kappa_{0}\) propagating at angle \(\theta\) to the positive \(x\) axis impinging on the permittivity profile of figure~\ref{rays}(c) scatters into as a function of angle. (c) The total reflected (\(R\)) and transmitted (\(T\)) intensities and their sum. It is only close to normal incidence that all of the wave is transmitted without diffraction. Specifically, over 99\(\%\) of the wave energy ends up in the \(T_{0}\) mode in the region \(0<\theta_{i}<\pi/60\).\label{varyangle}}
        \end{center}
    \end{figure}

Having considered the ability of this medium to suppress scattering for different frequencies of incidence, it is natural to also consider what happens when the angle of incidence deviates from normal incidence. Again there is no reason to expect transmission to be perfect at the design frequency at other angles of incidence, and this is indeed seen to be the case. The electric field norm is plotted for various angles of incidence at the design frequency \(ck_{0}=c\kappa_{0}\) in figure~\ref{diffractionpatternvaryangle}.
Away from normal incidence, a diffraction pattern is visible; the electric field norm has strong oscillations with a periodicity commensurate with that of the medium. This is indicative of other modes present in the solution for the field. Again we can get a quantitative description of the diffracted mode intensities. This is plotted in figure~\ref{varyangle}.
\footnotetext[1]{The ability to carry out such calculations based on the numerical simulations in Comsol is limited by the ability of the Perfectly Matched Layers (PMLs)~\cite{Berenger1994} used in the simulations at the boundaries of the unit cells to absorb any outgoing waves without reflection. The absorption rate decays exponentially with the wave vector component normal to the PML boundary. Therefore any of the simulations in which there are wave vector components close to grazing incidence will not be entirely diminished in the PML and will therefore introduce some errors. This manifests itself in having a total reflected and transmitted intensity summing to something other than unity. As such, angles (wave numbers) with modes close to grazing incidence have been removed from figure~\ref{varyangle}(c) (figure~\ref{varywavenumber}(c)).}

\section{TM Polarisation and the Geometrical Optics limit:}
Having designed a periodic medium exhibiting no diffraction to a TE polarised wave of a particular frequency, we now test its robustness to changing polarisation. Due to the symmetry in Maxwell's equations, the medium corresponding to having a permeability profile like that shown in figure~\ref{rays}(c) together with unit permittivity will not diffract a Transverse Magnetic (TM) polarised incident field. However, this need not be the case for the corresponding non-magnetic permittivity profile discussed earlier.
    \par
\begin{figure}[ht!]
        \begin{center}
	\includegraphics[width=\linewidth]{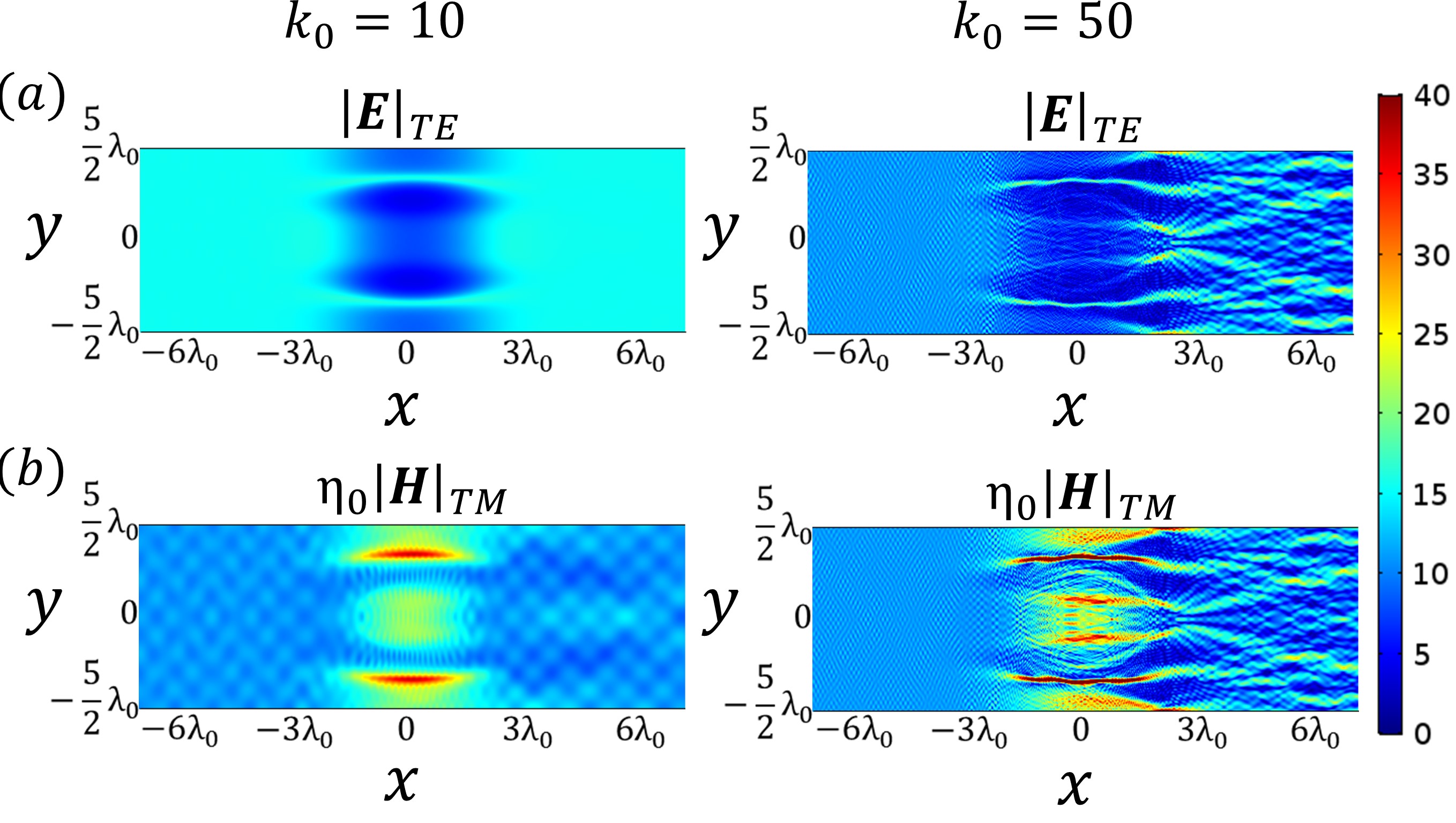}
        \caption{(a) The electric field norm \(|\textbf{E}|\) at wave numbers \(k_{0}=10\) and \(k_{0}=50\) for a TE polarised incident wave onto the medium described by the permittivity profile of figure~\ref{rays}(c), simulated  using Comsol Multiphysics~\cite{Comsol}. (b) The magnetic field norm multiplied by impedance of free space \(\eta_{0}|\textbf{H}|\) at the same wave numbers for a TM polarised incident wave.%(c) The absolute value of the difference between the two \(||\textbf{E}|-\eta_{0}|\textbf{H}||\). At \(k_{0}=50\), the diffraction pattern on either side of the medium matches quite closely.%
\label{polarisations}}
        \end{center}
    \end{figure}

The suppression of diffraction shown in the previous section hinges on the idea of being able to exactly map out rays in such a way that the energy flow is conserved (\(\nabla\cdot(A^{2}\nabla S)=0\)). However, if the incident field is instead Transverse Magnetic (TM) polarised, then the Helmholtz equation for the out-of-plane component of the magnetic field is modified to
    \begin{equation}
        \left[\nabla\cdot\left(\frac{1}{\epsilon}\nabla\right)+k_{0}^{2}\right]\varphi=0.\label{12}
    \end{equation}
As a result the second phase amplitude equation in for the field decoupled into amplitude and phase gets modified to
    \begin{equation}
        \nabla\cdot\left(\frac{A^{2}\nabla S}{\epsilon}\right)=0.\label{13}
    \end{equation}
for which the characteristic method now only solves for \(A/\sqrt{\epsilon}\). The remaining phase-amplitude equation is then a generalised version of the eikonal equation which is difficult to solve for the permittivity. It is only when one takes the geometrical optics limit, valid when \(k_{0}\gg|\nabla\epsilon|/\epsilon^{3/2}\), that the permittivity can be assumed locally homogeneous and thus for the conservation of energy equation (\ref{13}) to reduce to the more familiar \(\nabla\cdot(A^{2}\nabla S)=0\). The resulting fields for the two polarisations are compared in figure~\ref{polarisations}.
There is a visible diffraction pattern for the TM polarisation case at the wavenumber \(k_{0}=\kappa_{0}\) which diffraction is suppressed for TE polarisation. However, the difference between the polarisations diminishes as the wavenumber is increased. This can be understood from the first equation of (\ref{2}), which reduces to the eikonal equation \(\epsilon=(\nabla S)^{2}\) in the geometrical optics limit for both polarisations. With identical phase fronts, the factor of \(\epsilon\) in (\ref{13}) merely serves to rescale the field amplitude when going from TE polarisation to TM polarisation \(A\to\sqrt{\epsilon}A\). This leads to a greater amplitude of the TM polarised field inside the medium, without altering the diffraction pattern outside it.

\section{The Beam-shifter:}
As a second example we use our formalism to design a beam shifter. Beam-shifters have largely been designed using the coordinate transformations of transformation optics contained in~\cite{Pendry2006} using anisotropic media with graded permittivity and permeability tensors (see for example~\cite{Wang2008}). Such anisotropic profiles can be designed using metamaterials, for example using metallic rods~\cite{Salmasi2016} or tensor impedance surfaces~\cite{Patel2014}. Experimental realisations have so far been fairly limited but a structure based on transmission line metamaterials has been successful~\cite{Gok2013} and also in acoustics with perforated metamaterials~\cite{Wei2015}. All of these structures are based on transformation optics requiring anisotropic or magnetic materials. We instead propose an isotropic non-magnetic medium which laterally shifts a beam at a single frequency with negligible reflection.
    \par
So far, we have seen that the characteristic method has enabled the design of non-scattering permittivity profiles via a mapping out of the rays. Such numerical approaches are normally necessary to make progress due to the difficulty in solving PDEs exactly. However, there is a special case where the energy conservation equation can be solved exactly to give a permittivity profile with an interesting property, namely a wide beam is laterally shifted without reflection. We refer to this as a beam-shifter. We emphasise that this is not just geometrical optics; the beam-shifter we design is exact for wave optics.
    \par
\begin{figure}[ht!]
        \begin{center}
	\includegraphics[width=\linewidth]{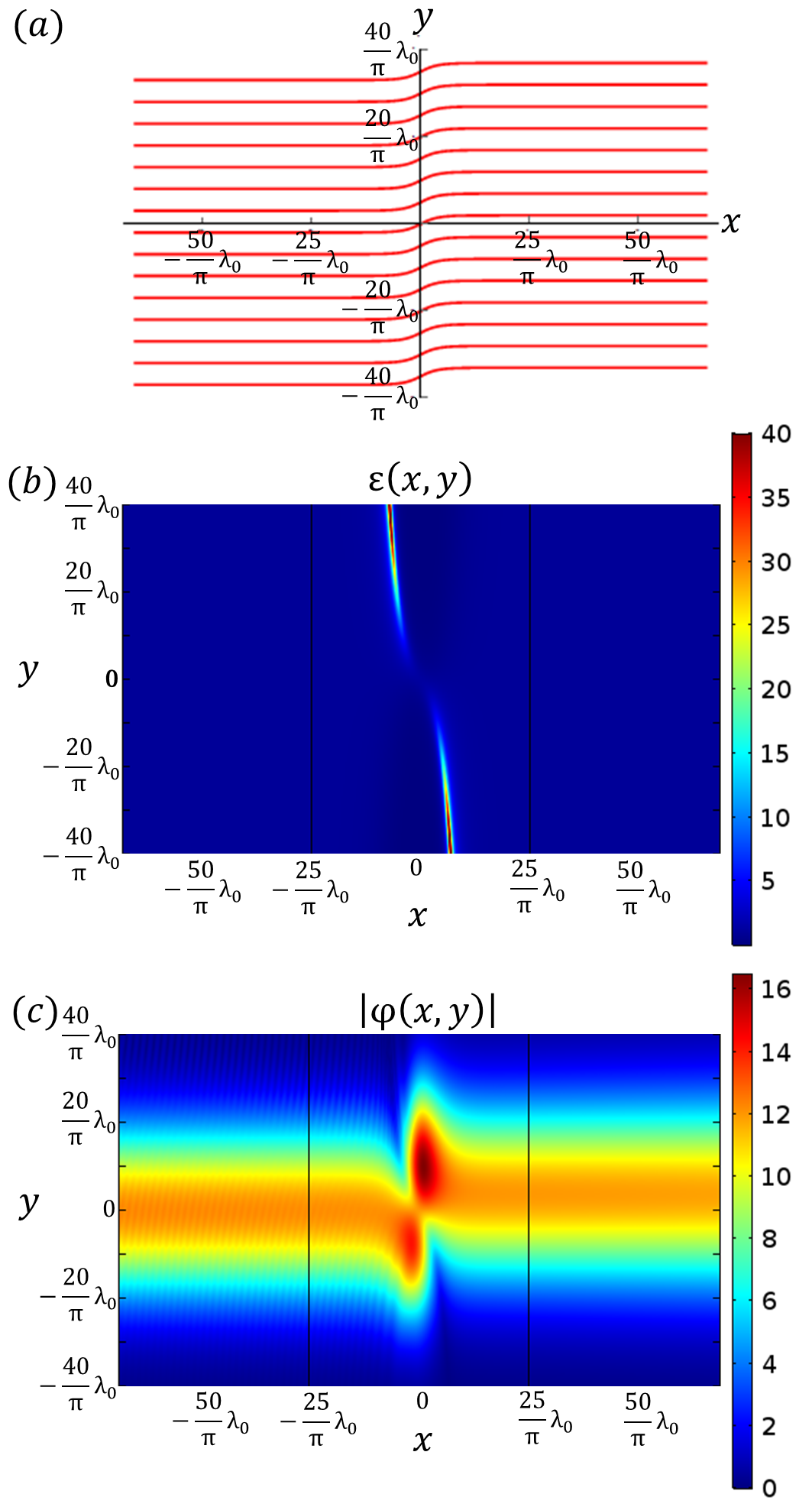}
        \caption{(a) The rays associated with the choices given in (\ref{21}) with \(\alpha=2\) and \(\beta=1\). (b) The corresponding permittivity profile with \(f=X^{-1}\) and \(k_{0}=\kappa_{0}\) in \(-25\lambda_{0}/\pi<x<25\lambda_{0}/\pi\) and free space either side. (c) The field norm corresponding to a right propagating incident Gaussian beam of width \(20\lambda_{0}/\pi\), simulated  using Comsol Multiphysics~\cite{Comsol}. The wave is transmitted with negligible reflection and with a beam shift of \(5\lambda_{0}/4\).\label{beamshifter}}
        \end{center}
    \end{figure}

Motivated by being able to solve the conservation of energy equation in one dimension (see equations (\ref{14}) and (\ref{15})), it is natural to solve the analogous two dimensional equation for a special case by imposing that (\ref{14}) holds for each of the individual coordinates. i.e.
    \begin{equation}
    \begin{split}
        \frac{\partial}{\partial x}\left(A^{2}\frac{\partial S}{\partial x}\right)&=0\\
        \frac{\partial}{\partial y}\left(A^{2}\frac{\partial S}{\partial y}\right)&=0.\label{16}
    \end{split}
    \end{equation}
which can be solved separately to give two expressions for the amplitude:
    \begin{equation}
        A=\frac{A_{y}(y)}{\sqrt{\frac{\partial S}{\partial x}}}=\frac{A_{x}(x)}{\sqrt{\frac{\partial S}{\partial y}}}.\label{17}
    \end{equation}
This can then be subsequently solved for the phase:
    \begin{equation}
        S=f(X(x)+Y(y)).\label{18}
    \end{equation}
where \(X'=1/A_{x}^{2}\) and \(Y'=1/A_{y}^{2}\). The particularly neat thing about this method of separating the equations for the different Cartesian coordinates is that the differential equation for the rays takes a separable form
    \begin{equation}
        \frac{dy}{dx}=\frac{Y'(y)}{X'(x)}.\label{19}
    \end{equation}
and, in particular, by taking \(Y(y)=y\), say, the slope of the rays depends only on the \(x\) coordinate and thus the rays are translationally invariant in the \(y\) direction. Meanwhile (\ref{2}) gives the expression for the permittivity
    \begin{equation}
    \begin{split}
        \epsilon&=(f')^{2}\left((X')^{2}+(Y')^{2}\right)\\
        \quad&+\frac{1}{2k_{0}^{2}}\left[\frac{X'''}{X'}+\frac{Y'''}{Y'}-\frac{3}{2}\left(\frac{(X'')^{2}}{(X')^{2}}+\frac{(Y'')^{2}}{(Y')^{2}}\right)\right]\\
        \quad&+\frac{1}{2k_{0}^{2}}\left[\left((X')^{2}+(Y')^{2}\right)\left(\frac{f'''}{f'}-\frac{3(f'')^{2}}{2(f')^{2}}\right)\right].\label{20}
    \end{split}
    \end{equation}
where only the first line would be retained in the geometrical optics limit. As for the periodic grating, our medium should sit in free space with a right propagating plane wave incident on the medium emerging totally as a right propagating plane wave without being scattered. To ensure that the rays are horizontal either side of the medium (\(X'\to+\infty\) as \(x\to\pm\infty\)), we choose, as a simple example,
    \begin{equation}
    \begin{split}
        X(x)&=\frac{\text{sinh}(\alpha x)}{\beta}\\
        Y(y)&=y.\label{21}
    \end{split}
    \end{equation}
leading to rays \(y=\frac{2\beta}{\alpha^{2}}\text{arctan}\left(\text{tanh}\left(\frac{\alpha x}{2}\right)\right)+\text{constant}\) which bend and straighten with a lateral shift of \(\pi/\alpha\), as shown in figure~\ref{beamshifter}(a).
To further ensure a right propagating plane wave either side of the medium, it is required that \(S\sim x\) as \(x\to\pm\infty\) so it is natural to choose \(f\) to be the inverse of \(X\):
    \begin{equation}
        f(z)=X^{-1}(z)=\frac{\text{arsinh}(\beta z)}{\alpha}.\label{22}
    \end{equation}
and we again choose a particular wavenumber of \(k_{0}=\kappa_{0}\) for the beam-shifter to function at. With these choices the permittivity profile obtained is shown in figure~\ref{beamshifter}(b) and is given by
    \begin{equation}
        \epsilon(x,y)=\frac{\beta^{2}+\alpha^{2}\text{cosh}^{2}(\alpha x)}{\alpha^{2}(1+(\beta y+\text{sinh}(\alpha x))^{2})}+\text{O}\left(\frac{1}{\kappa_{0}^{2}}\right).\label{23}
    \end{equation}
where the correction terms to the geometrical optics limit have been included in the plot but have been left out of (\ref{23}) for brevity. However, this is enough to explain the appearance of the permittivity profile. The denominator in (\ref{23}) reaches a minimum along \(y=-X(x)\). Along this channel the permittivity is higher than the surroundings in order to be able to bend the rays. In particular, as the channel's slope becomes increasingly vertical, the permittivity contrast needs to be greater in order for the rays to be bent by the same amount. As such, the permittivity in the channel increases quadratically in \(y\): \(\epsilon(X^{-1}(-y),y)=\text{O}(y^{2})\) as \(y\to\pm\infty\). Subsequently, this is likely to be difficult to realise practically (and indeed in simulations). However, with a wavelength of \(\lambda_{0}=2\pi/\kappa_{0}\) and a permittivity profile channel of similar width, it's possible to simulate this using a Gaussian beam with a width of a few wavelengths so that the contrast in the incident field has a negligible effect on the functionality of the beam-shifter. This also enables the lateral shift of the beam to become clear to see. The resulting shift in the beam can then be seen in a plot of the field norm, as shown in figure~\ref{beamshifter}(c). From the appearance of the permittivity profile only, we can explain how the rays are laterally shifted, but not why there is also no reflection from such a medium. However this is a general feature of graded index media; it is not easy to see why certain profiles with arbitrarily large contrasts are reflectionless, even in one dimension (such as the spatial Kramers-Kronig media~\cite{Horsley2016,Horsley2017} or the P{\"o}schl-Teller media~\cite{Epstein1930,Lekner2007}).

\section{Quantifying the effect of errors in the permittivity:}
The design procedure used in this work is exact. However, any errors in the permittivity, as seen in a realisation of the devices, will lead to errors in the corresponding field. Having already seen that the non-diffracting medium is surprisingly robust to slight changes in the wavenumber and the angle of incidence, it is hoped that the same might extend to slight changes in the profile. To quantify this, consider a slight perturbation to the permittivity, and the corresponding change to the field
    \begin{equation}
    \begin{split}
        \epsilon&\to\tilde{\epsilon}=\epsilon+\delta\epsilon\\
        \varphi&\to\tilde{\varphi}=\varphi+\delta\varphi.\label{24}
    \end{split}
    \end{equation}
The Helmholtz equation for the perturbed permittivity \((\nabla^{2}+k_{0}^{2}\tilde{\epsilon})\tilde{\varphi}=0\) can then be solved for small perturbations as
    \begin{equation}
        \delta\varphi(\textbf{x})=-k_{0}^{2}\int d\textbf{x}'G(\textbf{x}-\textbf{x}')\delta\epsilon(\textbf{x}')\varphi(\textbf{x}')\label{25},
    \end{equation}
where \(G(\textbf{x}-\textbf{x}')\) is the Green's function for the two-dimensional Helmholtz equation \((\nabla^{2}+k_{0}^{2}\epsilon(\textbf{x}))G(\textbf{x}-\textbf{x}')=\delta(\textbf{x}-\textbf{x}')\). In particular, the field response shows a linear dependence with the perturbed permittivity profile, so errors in the field can be made arbitrarily small by making the error in the permittivity arbitrarily small. In general, the errors in the permittivity, and in the reflection and the transmission can be quantified as
    \begin{equation}
    \begin{split}
        \eta_{\epsilon}&=\frac{1}{a^{2}}\int_{-6\lambda_{0}}^{6\lambda_{0}}dx\int_{-a/2}^{a/2}dy|\delta\epsilon(x,y)|\\
        \eta_{r}&=\frac{1}{a}\int_{-a/2}^{a/2}dy\frac{|\delta\varphi(-6\lambda_{0},y)|}{|\varphi(-6\lambda_{0},y)|}\\
        \eta_{t}&=\frac{1}{a}\int_{-a/2}^{a/2}dy\frac{|\delta\varphi(6\lambda_{0},y)|}{|\varphi(6\lambda_{0},y)|}\label{26},
    \end{split}
    \end{equation}
respectively, where \(a\) is the period of the grating. Using a Finite Difference Method (FDM) to calculate the Green's function, we calculated the error due to a simple discretisation of the profile in figure~\ref{rays}(c) into a \(20\) by \(20\) rectangular grid of homogeneous slabs. This corresponds to an error of \(\eta_{\epsilon}=0.604\) in the medium, and, using (\ref{26}), a fairly large error of \(\eta_{r}=0.372\) and \(\eta_{t}=0.191\) in the field. This is not surprising when we consider that it is the fine structure of the profile which leads to the removal of scattering---something that a simple \(20\) by \(20\) grid will not fully encapsulate. A finer \(200\) by \(200\) grid discretisation corresponds to an error \(\eta_{\epsilon}=0.0708\), and a very small error of \(\eta_{r}=0.0108\) and \(\eta_{t}=0.00614\) in the fields, which would not be noticeable in the field plots. When measuring the diffracted order energies (the squares of the fields), the errors will be significantly smaller---virtually all of the energy would be seen to transmit through the profile undiffracted. 

\section{Summary and Conclusions:}
We have constructed a a recipe for designing lossless media scatter-free media by choosing the exact rays to behave in a particular desired fashion and determining the corresponding amplitude and permittivity profile required for this whilst ensuring energy is conserved. This does not rely on the assumption of geometrical optics so is not just confined to the design of materials which vary on a scale much larger than the wavelength. We have applied the method to the design of non-scattering media whose spatial inhomogeneity is confined to a plane. Specifically we have designed a periodic graded-index permittivity profile with suppressed diffraction for a single frequency at normal incidence and a 'beam-shifter'; a graded-index permittivity profile which laterally shifts a Gaussian beam of a few wavelengths, without reflection. We expect that the method will also be useful for designing media in three dimensions which guide light in a desired fashion, such as 3d gratings, beam-benders and beam-expanders. Additionally this work raises the question of how well diffraction in periodic media can be controlled e.g. one could investigate the possibility of completely suppressing diffraction over a broadband frequency range or for a range of angles of incidence. Also, one could look at whether or not it is possible to diffract perfectly into a pair of modes instead of just one.

\section{Acknowledgements:}
CGK acknowledges financial support from the EPSRC Centre for Doctoral Training in Electromagnetic Metamaterials EP/L015331/1. SARH acknowledges financial support from the Royal Society and TATA (RPG-2016-186).

\end{document}